\documentstyle[aps,prb,multicol,bezier]{revtex}

\begin{document}
\title{
   Crossover from Selberg's type to Ruelle's type zeta function
   in classical kinetics.
}

\author{ Daniel L. Miller }
\address{
 Dept. of Physics of Complex Systems,\\
 The Weizmann Institute of science,
 Rehovot, 76100 Israel                \\
 e-mail  fndaniil@wicc.weizmann.ac.il
}

\date{\today}
\maketitle

\begin{abstract}
   The decay rates of the density-density correlation function are computed
   for a chaotic billiard with some amount of disorder inside. In the case of
   the clean system
   the rates are zeros of Ruelle's Zeta function and in the 
   limit of strong disorder they are roots of Selberg's Zeta function. 
   We constructed
   the interpolation formula between two limiting Zeta functions by analogy
   with the case of the integrable billiards. The almost clean limit is
   discussed in some detail.\\  
   \\                              
   PACS numbers: 05.20.Dd, 05.45.+b, 51.10.+y
\end{abstract}
\pacs{ 05.20.Dd, 05.45.+b, 51.10.+y}

\begin{multicols}{2}

\narrowtext

It is natural to assume that chaotic billiards with a small amount of
disorder\cite{Atland-Gefen-95,Agam-Fishman-96,Aleiner-Larkin-96} are good
models for ballistic cavities which have been employed in a number of recent
experiments, see Ref.\onlinecite{Marcus-97}. Such a model is interesting
because the disorder (in two dimensions) can be characterized by one
parameter: the elastic scattering time $\tau$.  The mixing properties of this
model in  two important limits  $\tau\rightarrow0$ and
$\tau\rightarrow\infty$ were known in the literature, see
Refs.\onlinecite{Landau-par19,Bunimovich-oct85}. In the present work we
discuss the crossover from one limit to the other for some two-dimensional
billiard.   The case of the three-dimensional billiards is more complicated.
Particularly, the uniform scattering in three dimensions leads to very fast
resonant mixing. At the end of the paper we provide the generalization of our
results for the case of three dimensions.

Let us focus attention on the eigenvalues and the eigenmodes of the
kinetic equation for the distribution function $f(\vec r,\phi)$ of 
non-interacting particles inside a two-dimensional billiard. This function
is defined on the constant energy manifold $|\vec v_\phi|=v=$const, and $\vec
v_\phi = (v\cos(\phi), v\sin(\phi))$.  The precise form of the kinetic equation
depends on the details of the impurity potential, but we are going to
investigate two models
\begin{equation}
   {\partial f \over \partial t}
   + \vec v_\phi\vec \nabla f =
   \left\{\begin{array}{ll}
      {\bar f - f \over \tau} \;, &\;\;\text{model 1}\\
      {1\over \tau}
      { \partial ^2 f \over \partial \phi^2} \;, &\;\;\text{model 2}
   \end{array}\right.\;,
\label{eq:intr.1}
\end{equation}
where $\bar f(\vec r) = \int\,f(\vec r, \phi)\,d\phi/(2\pi)$. The above
equation has to be solved with mirror boundary conditions   $f(\vec r,\phi) =
f(\vec r, 2\alpha(\vec r) -\pi - \phi)$, where $\vec r$ is taken on the boundary
of the billiard and $\vec n = \Bigl( \cos(\alpha), \sin(\alpha) \Bigr)$ is
normal to the boundary. Equation~(\ref{eq:intr.1}) has a special solution
$f_0(\vec r, \phi)=$ const
for all values of $\tau$ and we will ignore it in the rest of the paper.

In both models, the collision integral conserves energy. The first model
corresponds to uniform scattering  in all directions and the second model
is valid if small angle scattering is dominant. Let us look for solutions
proportional to $e^{-s_nt}$. The eigenvalues $s_n$ of the kinetic equation
are so-called mixing rates, or decay rates of the density-density correlation
function when $\tau\rightarrow0$ or Ruelle's resonances when
$\tau\rightarrow\infty$.

These resonances can be found as zeros of the spectral determinant $Z(s)$.
Let us start to compute $Z(s)$ in the limit of pure chaos. Cvitanovic and
Eckardt\cite{Cvitanovic-Eckardt-91} have computed the spectral determinant
for the axiom A system, but the result is the expansion over the unstable
periodic orbits and it seems to be valid for wide class of systems. Therefore
in the limit $\tau\rightarrow\infty$ the spectral determinant is
\begin{equation}
  -\log(Z(s)) = \sum_{p} \sum_{r=1}^\infty
  {1\over r}
  {1 \over |\text{det}(I-M_p^r)|}
  e^{sl_pr/v}\;,
\label{eq:intr.2}
\end{equation}
where $v$ is the velocity.
This expression contains the sum over the primitive periodic orbits $p$ taken
with repetition $r$. In the case of billiards, the action for the periodic
trajectory is just the length $l_pr$. Each oscillating term in the sum in
Eq.~(\ref{eq:intr.2}) is weighted by the stability amplitude which behaves on
average as
\begin{equation}
  {1 \over |\text{det}(I-M_p^r)|}
  \equiv e^{-\lambda_{pr} l_p r/v}
  \approx e^{-\lambda l_p r/v}
  \;,
\label{eq:intr.2a}
\end{equation}
where the first equality defines $\lambda_{pr}$, and
$\lambda$ is the Lyapunov exponent of the billiard. The
Zeta function given by Eq.~(\ref{eq:intr.2}) is of Ruelle's type.

When the disorder is strong, the kinetic equation can be transformed into the
diffusion equation for $\bar f$
\begin{equation}
   {\partial \bar f \over \partial t}
   - {v^2\tau\over 2}
   \nabla^2 \bar f = 0\;,
\label{eq:intr.3}
\end{equation}
which has to be solved with boundary conditions $\vec n\nabla \bar f = 0$.
This equation allows one to find the decay of modes with $\bar f\not\equiv0$.
The decay rates for modes with $\bar f=0$ for all $\vec r$ should be computed
in a different way.

In the limit $\tau\rightarrow0$ we can use the ``semiclassical'' approximation
for equation
\begin{equation}
   (k^2+\nabla^2) \bar f = 0 \;,
\label{eq:intr.4}
\end{equation}
where $k^2 = 2s/(v^2\tau)$. The logarithm of the Selberg's type Zeta function
is again the sum over periodic orbits\cite{Gutzwiller-book91,Voros-88}
\begin{equation}
  -\log(Z(s)) = \sum_{p} \sum_{r=1}^\infty
  {1\over r}
  {1 \over \sqrt{|\text{det}(I-M_p^r)|}}
  e^{ikl_pr}\;.
\label{eq:intr.5}
\end{equation}
Here the phase factors correspond to the case when  Eq.~(\ref{eq:intr.4})
should be solved with the Neumann boundary condition and Maslov's indexes 
vanish.\cite{MartinSieber-95}
The natural question to ask is whether it is
possible to compute the decay rates of the modes with $\bar f\not\equiv0$
for all values of $\tau$ by constructing a suitable Zeta function?

We can understand the connection between different types of Zeta functions
by making use of the following approximation
\begin{eqnarray}
   -\log(A(x)) &\equiv& \sum_{p} e^{x l_p}
\\
   Z(s) &\approx& A({s-\lambda \over v}) \;\;\;\;\tau\rightarrow \infty
\label{eq:comm.1}
\\  
   Z(s)  &\approx&  A(i\sqrt{2s\over v^2\tau} -
   {\lambda\over 2v } ) \;\;\;\;\tau\rightarrow 0
\label{eq:comm.2}
\end{eqnarray}
where we have neglected the repetitions of periodic orbits and fluctuations of
the Lyapunov exponent in Eqs.~(\ref{eq:intr.2}) and (\ref{eq:intr.5}). 
Let us denote the eigenvalues of the wavenumber in  Eq.~({\ref{eq:intr.4}}) 
as $q_n$,
$n=1\ldots\infty$. We  therefore suggest that eigenvalues of the kinetic
equation are
\begin{equation}
   s_n =
   \left\{\begin{array}{ll}
      q_n^2v^2\tau/2\,&\;\;\tau\rightarrow0\\
      \lambda/2 + i q_n v\,&\;\;\tau\rightarrow\infty
   \end{array}\right.\;,
\label{eq:intr.6}
\end{equation}
where $\lambda$ is the mean Lyapunov exponent. In order to obtain the  second
expression we have noted that the function $A(x)$ should have zeros $x_n=iq_n
- {\lambda\over 2v }$ and we have made use of Eq.~(\ref{eq:comm.1}). We would
like to emphasize that Eq.~(\ref{eq:intr.6}) has nothing to do with the
convergence properties of the Zeta functions. It based on the fact that
Ruelle's type Zeta function Eq.~(\ref{eq:intr.2}) and Selberg's type Zeta
function are almost the same function $A(x)$ taken with different arguments.
The second case in  Eq.~(\ref{eq:intr.6}) means that there is a shift of 
$\lambda/2$ between the zeros of Ruelle's type Zeta function and the zeros of
Selberg's type Zeta function. Such a shift was observed in the problem of
quantum and classical scattering in a three disk problem, compare Figs. 2.14
and 3.6 of  Ref.~\onlinecite{Gaspard-resonances} and see
Refs.\onlinecite{Gaspard-Rice-88,Gaspard-Ramirez-92} for details. The main
idea of this work is to obtain known Zeta functions as the two
limits of one kinetic Zeta function.

\begin{figure}
\unitlength=1mm
\linethickness{0.4pt}
\begin{picture}(83.00,83.00)
\put(6.00,6.00){\framebox(77.00,77.00)[cc]{}}
\bezier{96}(20.00,6.00)(20.00,20.00)(10.00,20.00)
\bezier{256}(40.00,6.00)(40.00,40.00)(10.00,40.00)
\bezier{496}(70.00,6.00)(70.00,70.00)(10.00,70.00)
\put(9.00,6.00){\vector(1,0){1.00}}
\put(29.00,6.00){\vector(1,0){1.00}}
\put(53.00,6.00){\vector(1,0){1.00}}
\put(11.00,70.00){\vector(-1,0){1.00}}
\put(11.00,40.00){\vector(-1,0){1.00}}
\put(11.00,20.00){\vector(-1,0){1.00}}
\put(20.00,6.00){\vector(0,1){1.00}}
\put(40.00,6.00){\vector(0,1){1.00}}
\put(70.00,6.00){\vector(0,1){1.00}}
\put(56.00,53.00){\vector(-1,1){0.99}}
\put(33.00,31.00){\vector(-1,1){1.00}}
\put(16.60,17.55){\vector(-1,1){0.48}}
\put(43.00,3.00){\makebox(0,0)[cc]{Re$s$}}
\put(3.00,43.00){\makebox(0,0)[cc]{Im$s$}}
\put(10.00,3.00){\makebox(0,0)[cc]{$\lambda/2$}}
\end{picture}
   \caption{The decay rates of the density-density correlation function
   move on the complex plane when disorder decreases, as shown by arrows.
   The termination point is Re$s\approx\lambda/2$, where $\lambda$ is Lyapunov
   exponent. }
\label{fig:roots}
\end{figure}
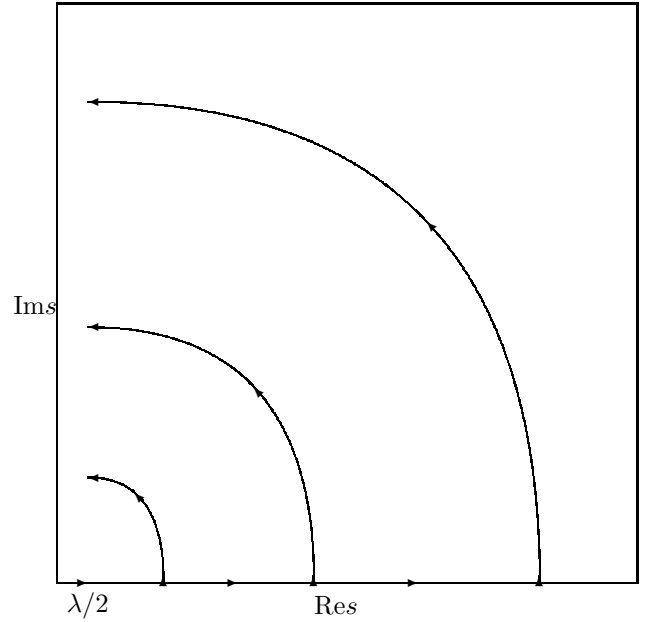

Let's examine first the integrable case. Model 1 for the square billiard
was solved by Atland and Gefen\cite{Atland-Gefen-95}, and Agam and
Fishman\cite{Agam-Fishman-96}, who modeled the short range random potential
by random spheres or circles. For the square billiard of size $L$ the
spatial dependence of the density $\bar f(r)$ is $\sum\exp(i\vec q\vec r)$,
where $q$ is such that $\sin(q_xL)\sin(q_yL)=0$, and the sum is over four
possible directions of $\vec q$. Then the values of $q$ are ``quantized'',
and we  will denote them $q_n$, and the modes with $\bar f\not\equiv0$ can be
numbered
\begin{equation}
    f_n(\vec r,\phi,t) \propto \sum { \tau^{-1}\over -s + \tau^{-1} +
    i\vec v_{\phi}\vec q_n + 0}
    e^{i\vec q_n \vec r-st}\;,
\label{eq:square.1}
\end{equation}
where the sum is over four
possible directions of $\vec q_n$. Integration over $\phi$
leads to the equation for $s_n$
\begin{equation}
    \tau^{-2} = ( s_n - \tau^{-1} )^2 + v^2 q_n^2\;,
\label{eq:square.2}
\end{equation}
and the corresponding Zeta function is
\begin{equation}
  -\log(Z(s)) = \sum_{p} 
  \sqrt{L^4q\over \pi^{3}l_p^{3}}
  e^{ iql_p}\;,
\label{eq:square.3}
\end{equation}
where $p$ is not a single orbit but the resonant tori
\cite{Berry-feb85},
and the connection between $s$ and $q$ is as in Eq.~(\ref{eq:square.2}).

In the case of model 2, the solutions are still proportional to
$e^{i\vec q\vec r}$, but the angular dependence is different. The solution
with $\bar f\not\equiv0$ is the ``ground state'' of
\begin{equation}
   \left[-s + i\vec v_\phi \vec q_n
   - {1\over \tau} {\partial^2\over \partial \phi^2}\right]f_n =0\;,
\label{eq:square.4}
\end{equation}
because the real parts of the decay rates are positive. Surprisingly,
Eq.~(\ref{eq:square.2}) gives a numerically good approximation for $s_n$
for this model. Other ``angular'' modes, which have $\bar f=0$, are very
different for models 1 and 2.

It is not easy to compute eigenmodes of Eq.~(\ref{eq:intr.1}) for the
integrable billiards which have other than rectangular shapes. For such cases
Eq.~(\ref{eq:square.3}) becomes an interpolation formula for the kinetic Zeta
function. One should only replace the pre-exponential factor by the amplitude
from the Berry -- Tabor\cite{Berry-Tabor-mar77} formula. For example, the
resonant tori for the circular billiard of radius $R$ are numbered by
the winding number $M$ and by the number of vertices $n$, have length
$l_{Mn}= nL_{Mn}/\pi$, where $L_{Mn}= 2\pi R\sin(\pi M/n)$. Then one can use
Eq.~(\ref{eq:square.3}) after the replacement $p\rightarrow Mn$, and 
$L\rightarrow L_{Mn}$, see Ref.\onlinecite{Smilansky-rev95}.

\begin{mathletters}
Combining together Eqs.~(\ref{eq:intr.2}), (\ref{eq:intr.5}),
(\ref{eq:square.2}), and (\ref{eq:square.3}) we can introduce the kinetic Zeta
function as
\begin{equation}
  -\log(Z(s)) = \sum_{p} \sum_{r=1}^\infty
  {1\over r}
  e^{ [\sqrt{( s-2/\tau_{pr})s}-\lambda_{pr}/2 ]\,l_pr/v } 
  \;,
\label{eq:new.5}
\end{equation}
where
\begin{equation}
  \tau_{pr} =
  \left\{\begin{array}{lr}
     \tau,&  \lambda_{pr} \tau/2 < 1 \\
     2/\lambda_{pr}, & \lambda_{pr} \tau/2 > 1
     \end{array}\right.\;.
\label{eq:new.6}
\end{equation}
For $\lambda\tau<1/2$ the kinetic Zeta function coincides with Selberg's type
Zeta function Eq.~(\ref{eq:intr.5}) in the domain of the complex $s$ plane
$|s\tau|\ll 1$. For $\lambda\tau>1/2$ the kinetic Zeta function becomes
independent of $\tau$ and coincides with Ruelle's type Zeta function
Eq.~(\ref{eq:intr.2}) in the domain of the complex $s$ plane $|s\tau|\gg 1$.
\label{eq:new.7}\end{mathletters}

The interpolation formula Eqs.~(\ref{eq:new.7}) for the kinetic Zeta
function implies the following interpolation formula for the decay rates
\begin{equation}
   s_n =
  \left\{\begin{array}{lr}
     \displaystyle
     {1\over\tau} - \sqrt{{1\over\tau^2}- (q_nv)^2} , 
     & {1\over q_nv} \ge \tau  \\
     \displaystyle
     {1\over\tau} + i \sqrt{ (q_nv )^2- {1\over\tau^2}} , 
     & {2\over\lambda} \ge \tau \ge { 1\over  q_nv }\\
     \displaystyle
     {\lambda\over2} + iq_nv, &  \tau \ge  {2\over\lambda}
  \end{array}\right.
\label{eq:new.8}
\end{equation}
where $q_n$ are the eigenvalues of the wavenumber in Eq.~({\ref{eq:intr.4}}).
There is a gap between the last two expressions
$s_n|_{\tau=2/\lambda-0}-s_n|_{\tau=2/\lambda+0}\sim \lambda^2/(q_nv)$, which
is numerically small for most cases. The motion of the decay rates on the
complex plane is schematically shown in Fig.~\ref{fig:roots}. In the limit of
strong disorder some of the $s_n$ are on the real axis, and the imaginary part of
$s_n$ becomes non-zero when $\tau q_n v = 1$. Then $s_n$ moves along the arc
and stops when $\tau = \lambda/2$.

Equations~(\ref{eq:intr.6}) and (\ref{eq:new.8}) show that a chaotic system is
qualitatively  different from a diffusive system from the point of view of
the position of Ruelle's resonances $s_n$ on the complex plane. In the
chaotic limit all resonances lie on a line parallel to the imaginary
axis. The disorder induces motion of the resonances toward the real axis
as was found by Agam and Fishman.\cite{Agam-Fishman-96}

The interpolation formula between Ruelle's type and Selberg's type Zeta
exists only if the diffusion modes transform to the so-called
Frobenius -- Perron modes as the disorder goes to zero. This has not yet been
proven for our case.  The main difficulty is that the diffusion modes are selected from
all kinetic modes by the condition $\bar f\not\equiv0$. At the same time
Frobenius-Perron modes are selected by the choice of the functional space.
However, in other systems one can consider the diffusion
modes as modes of the Frobenius -- Perron operator.\cite{Gaspard-96}

Some additional information might be obtained from the properties of the
propagator of Eq.~(\ref{eq:intr.1}), which can be written as a path integral
for model 2
\begin{eqnarray}
  && G(\vec r, \phi, \vec r_0, \phi_0, t) =
  \int
  D[\psi]
  \delta(\vec r-\vec r_0 - \int_0^t \vec v_{\psi} dt)
\nonumber\\
   &&\times
  e^{-{\tau\over 4}\left\{\int_0^{t_1} \dot \psi^2 dt
  +\int_{t_1}^{t_2} \dot \psi^2 dt+\ldots+\int_{t_n}^{t} \dot \psi^2
  dt\right\}}
\label{eq:path.1}
\end{eqnarray}
where $\psi(0)=\phi_0$, $\psi(t_j-0)+\psi(t_j+0)=2\alpha_j$, ... , 
$\psi(t)=\phi$, and the path $\vec r_0 + \int_0^{t} \vec v_{\psi} dt$ 
touches the boundaries $n$ times at the points $\vec r_1,\ldots,\vec r_n$, at
the times $t_1,\ldots,t_n$. The angle  $\alpha_j$ is the direction of
the tangent to the boundary at the reflection point $\vec r_j$. The trace of
the propagator Eq.~(\ref{eq:path.1}) known also as the return probability is
\begin{equation}
  p(t)=
  \int d\vec r_0
  \int d\phi \int D[\psi]
  e^{-{\tau\over 4}\int_0^t \dot \psi^2 dt}
  \delta(\int_0^t \vec v_{\psi} dt) \;,
\label{eq:path.2}
\end{equation}
where $\psi(t)= \psi(0)=\phi$ and $\int_0^t \dot \psi^2 dt$ is defined as in 
Eq.~(\ref{eq:path.1}).

The propagator Eq.~(\ref{eq:path.1}) should interpolate between the Frobenius
-- Perron operator in the limit $\tau\rightarrow\infty$ and  the diffusion
operator in the limit $\tau\rightarrow0$. Then the trace of this propagator
Eq.~(\ref{eq:path.2}) should provide us with a systematic  way to compute the
interpolation formula for the Zeta function, because $-\log\,Z(s) = \int_0^\infty
e^{st}\,t^{-1} p(t)dt$. Here the sign of $st$ in the Laplace transform is
positive, because we want the roots of the Zeta function to have  the meaning of the decay
rates.

In the limit of weak disorder $\tau\rightarrow\infty$, one may hope to
obtain the small corrections $\propto {1\over\tau}$ to the Frobenius -- Perron
operator, and therefore to Eq.~(\ref{eq:intr.2}). Particularly one may expect
to obtain the additional ``stabilization'' of the periodic orbits through the
disorder. Let's consider the vicinity of the periodic orbit $p$ in phase
space. The path in such a vicinity can be described by the coordinate $x(t)=vt$
along the orbit, by the coordinate $y(t)$ normal to the orbit, and by the
deviation of the direction of motion $\phi(t)$. The position of the particle
at the end of the path and at the beginning of the path are connected by
\begin{equation}
  \left(\begin{array}{c} y(t) \\ \phi(t) \end{array}\right)
  =
  M_p
  \left(\begin{array}{c} y(0) \\ \phi(0) \end{array}\right)
  + \sum_{j=1}^{n_p} M_{pj}
  \left(\begin{array}{c} \theta_j {L_{pj}\over 2} 
  \\ \theta_j \end{array}\right)
  \;,
\label{eq:path.3}
\end{equation}
where the orbit $p$ crosses the billiard $n_p$ times. In other words the
orbit consists of $n_p$ segments of length $L_{pj}$. When the particle is
going along the segment $j$, it can be scattered by the disorder at small
angle $\theta_j$, and then the rest of the path is distorted too. The
cumulative change of the end of the path is given by the sum in the right
hand side of Eq.~(\ref{eq:path.3}), where $M_{pj}$ is the monodromy matrix of
the piece of the orbit consisting of the segments $L_{pj+1},\ldots,
L_{pn_p}$. One can see immediately from Eq.~(\ref{eq:path.3}), that the
stability amplitude of the closed path $y(t)=y(0)$, $\phi(t)=\phi(0)$ is
independent of $\theta_1, \ldots, \theta_{n_p}$, and therefore it is
independent of $\tau$. Therefore, there are no $1/\tau$ corrections to the
Zeta function Eq.~(\ref{eq:intr.2}) and our interpolation formula
Eq.~(\ref{eq:new.7}) is independent of $\tau$ for $\tau> 2/\lambda$.

In the case of the three dimensional billiards the effect of the disorder is
different because the scattering becomes three-dimensional and the
distribution function depends on the three coordinates and two angles.
For the cubic billiard the spatial dependence of the density is again $\sum
e^{i{\vec q}{\vec r}}$, where the sum is
over the six orthogonal orientations of ${\vec q}$, and the modes are selected by the
rule $\sin(q_xL) \sin(q_yL)\sin(q_zL)=0$. Then, the dispersion relation for the
analog of model 1, (uniform scattering), can be found in
Ref.\onlinecite{Morse-Feshbach-book}:
\begin{equation}
   1 - s\tau = {qv\tau \over \tan(qv\tau) }\;,
\label{eq:3D.1}
\end{equation}
where $qv\tau<\pi$. In other words there are no modes with $qv\tau\ge\pi$,
and $\bar f\not\equiv0$, where the bar means the average over the solid
angle. Equation~(\ref{eq:3D.1}) describes the diffusion modes for small
$\tau$, but  it cannot be used for large $\tau$. If the mode has $q$
close to $\pi/(v\tau)$ then the decay of such a mode is very fast
$s\propto(\pi-qv\tau)^{-1}$.

The model with small angle scattering in three dimensions is the Fokker --
Planck equation for the distribution function, which should be solved together
with mirror boundary conditions on the billiard boundary. The solutions inside
the cubic billiard have the same dispersion $s(q)$ as in the case of the
square billiard, if $\bar f\not\equiv0$.  Therefore one may hope that
Eq.~(\ref{eq:new.7}) gives the interpolation of the Zeta function of
the Fokker -- Planck equation for modes with $\bar f\not\equiv0$.

In summary, we have constructed the interpolation formula for the Zeta
function of the kinetic equation, in both ``chaotic'', Eqs.~(\ref{eq:new.7}),
and ``integrable'', Eq.~(\ref{eq:square.3}) cases. From the mathematical
point of view our kinetic Zeta function interpolates between Ruelle's 
and Selberg's  Zeta functions. Our formulas are
independent of the particular choice of the collision integral for
two-dimensional billiards and are suitable for small angle scattering in three
dimensions.

\acknowledgments
It is my pleasure to thank prof. Smilansky for important remarks and
discussions.
This work was supported by Israel Science Foundation and the Minerva Center
for Nonlinear Physics of Complex systems.


\end{multicols}
\end{document}